# Application of Monte Carlo Simulations

# to Improve Basketball Shooting Strategy

**Byeong June Min**

*Department of Physics, Daegu University, Kyungsan 712-714, Korea*

The underlying physics of basketball shooting seems to be a straightforward example of the Newtonian mechanics that can easily be traced by numerical methods. However, a human basketball player does not make use of all the possible basketball trajectories. Instead, a basketball player will build up a database of successful shots and select the trajectory that has the greatest tolerance to small variations of the real world.

We simulate the basketball player's shooting training as a Monte Carlo sequence to build optimal shooting strategies, such as the launch speed and angle of the basketball, and whether to take a direct shot or a bank shot, as a function of the player's court positions and height. The phase space volume $\Omega$ that belongs to the successful launch velocities generated by Monte Carlo simulations are then used as the criterion to optimize a shooting strategy that incorporates not only mechanical, but human factors as well.

Email: bjmin@daegu.ac.kr

Fax: +82-53-850-6439, Tel: +82-53-850-6436

## I. INTRODUCTION

It may seem that the underlying physics of basketball is too well understood to merit another survey. That may well be the reason why there is so little research into the physics of basketball, despite the huge popularity of the game. Furthermore, we possess enough computational power to realistically trace the motion of a basketball. These circumstances seem to suggest that we only need to calculate all the physical trajectories of a basketball and run the appropriate statistical analysis. This has been the approach of earlier researchers who studied the physics of basketball [1-7].

However, it is not possible for a human player to make use of all the possible basketball trajectories, nor is it solely on account of human shortcomings. High launch angle shots occupy less phase space volume because the velocity space volume element is proportional to the cosine of the launch angle measured from the horizontal plane. Smaller phase space volume means less tolerance for small inaccuracies in aim. Such an attempt would have a great entertainment value, but nobody would try it under a pressure to win.

We will introduce the phase space volume associated with the launch velocity as an important criterion in the optimization of the shooting strategy. We consider a basketball player shooting from a position in the court, given by $(x, y)$, where $x, y = 1, 2, 3\, m$ defined relative to the center of the hoop (Figure 1). We consider two shooters with different heights.

The successful throws have usually been analyzed in the form of the launch angle $\theta$ vs. the launch speed $v$ diagram. The launch angle is measured from the horizontal plane, following the ordinary usage of the word and the conventions in the previous studies. We should keep in mind that the simple $\theta - v$ diagram cannot faithfully represent the three-dimensional velocity phase space.

The details of $\theta - v$ diagram may vary as a function of the shooter's position from the hoop, but, at an intermediate distance and angle, generally take a horseshoe shape. The $\theta - v$ diagram for all the bank shots that can be made by a shooter located at $(3,3)\, m$ from the center of the hoop with a release height of $1.9\,m$ is shown in Figure 2. In this calculation, we scanned the velocity phase space thoroughly, using $\Delta v = 0.1\,m/s$, $\Delta\theta = 1°$, and $\Delta\varphi = 1°$.

It is not possible to determine the optimal $\theta_0$ and $v_0$ without further assumptions. It is also clear that any *ad hoc* scheme to select a portion of the $\theta - v$ diagram will also affect the optimal values $\theta_0$ and $v_0$.

We analyze the bank shots in Figure 2 by replacing each bouncing spots on the backboard with a Gaussian function. Then, the bouncing spots will become a continuous distribution on the backboard. After normalization, it should correspond to the probability density of bank shots on the backboard. We observe that the distribution is elongated upward (Figure 3). In Figure 3, only the right half of the backboard above the rim height is shown. The horizontal axis should correspond to the x-axis, and the vertical axis to the z-axis. This probability distribution is counter-intuitive in that it stretches too high on the backboard.

If we consider how a person learns to shoot, a basketball player will try a shot and then judge his own shooting by how far he has missed the hoop. Based on the deviations from the goal, he will try to improve his aim by making adjustments. Eventually, he will find optimal $\theta_0$ and $v_0$ as a function of his court position and his height through a lengthy shooting practice. To do that, he will have to choose a part of the horseshoe shape, because the diagram extends to unlikely $\theta$ and $v$. But, a straightforward application of the deterministic Newtonian equations alone cannot offer a criterion for such a selection process.

Tran *et al* [4] introduce standard deviations $\sigma_\theta$ and $\sigma_v$ about the optimal values of $\theta$ and $v$, and keep the successful shooting ratio at a value typical of professional players. The skill level of the player is reflected in the successful shooting ratio. This scheme would be equivalent to selecting a rectangular region in the $\theta - v$ diagram centered at the $\theta_0$ and $v_0$. There exist arbitrariness in setting the optimal values ($\theta_0$ and $v_0$) and the standard deviations ($\sigma_\theta$ and $\sigma_v$).

We propose to view the basketball player's shooting training process as a Monte Carlo sequence. A basketball player will judge his shootings by how far the basketball is located from the center of the rim when it passes the horizontal plane located at the height of the hoop. The skill of the basketball player will be described by the temperature parameter of the Monte Carlo simulation. There is no need for additional parameters in this approach.

As Silverberg *et al* [5] point out, the players tend to shoot more accurately when they are closer to the hoop. The ability of the shooter to control the launch parameters does not change as a function of his position. But, the nearer to the hoop the shooter is positioned, the more launch angle and speed (thus, the more phase space volume) is available to the shooter. Therefore, we choose to calculate the phase space volume occupied by successful launch velocities from the Monte Carlo sequence and take it as the criterion for the better strategy.

## II. NUMERICAL MODEL

The dynamics of the basketball is governed by gravitational force and air resistance. The effect of the Magnus force for basketball is ignored because of its small magnitude. However, we assumed that the basketball is shot with a back spin and that the back spin results in a gain in the downward velocity as it bounces off the backboard, given by $\Delta v \, \hat{n} \times (\hat{v} \times \hat{z})$, where $\hat{n}$ is the unit vector normal to the backboard and $\Delta v = 0.2 \, m/s$. The basketball is assumed to be a rigid thin shell with a restitution coefficient of 0.75. The air resistance force is expressed by Okubo *et al* [6,7] as

$$\vec{F}_{air} = -\frac{1}{2} C_d \rho A v \vec{v},$$

where $\rho$ is the density of air, $C_d = 0.54$ the drag coefficient of the basketball in air, $A$ the cross-section area of the basketball, and $\vec{v}$ is the velocity of the basketball. The equations were integrated using the fourth-order Runge-Kutta method with a time step of $1\times 10^{-3}\,s$.

In the Monte Carlo simulation, the key variable is the horizontal deviation of the basketball from the center of the hoop $R$. A new launch velocity will be accepted if it passes nearer to the center of the hoop. If $R$ is larger, the new launch velocity will be accepted with a probability of $\exp(-\Delta R/\kappa)$. We used $\kappa = 0.5\,m$ throughout the Monte Carlo simulations. A smaller value will correspond to a more accurate shooter.

To calculate the phase space volume $\Omega$ occupied by the successful shots within the Monte Carlo sequence, we set up a grid in the $\vec{v}$-space at an interval of $0.01\,m/s$ and counted the number of cubes that belonged to the successful shots. The smaller the phase space volume $\Omega$, the more difficult the shot will be.

We compared bank shots and direct shots made from rectangular grid positions on the court. We define the z-axis as the upward direction, the x-axis starting at the center of the hoop and stretching toward the sideline being parallel to the backboard, and the y-axis stretching toward the center line being perpendicular to the backboard (Figure 1). We considered two release heights $z = 1.9\,m$ and $z = 2.1\,m$ to investigate how the player's height influences the output.

## III. RESULTS AND DISCUSSION

Figure 4 and Figure 5 depict the $\theta - v$ diagram and the bank shot distributions on the backboard, respectively, when the shooter is located at $(3,3)\,m$ from the center of the hoop with a release height of $z = 1.9\,m$. In this case, the Monte Carlo simulation produces a range of shooting parameters consistent with our intuition. The $\theta - v$ diagram does not extend to extreme values, but remain in the neighborhood of intuitive values.

We can see that the backboard profile is now more symmetrical and does not unduly extend upwards. However, its center has moved slightly upward, which means that the undue contribution from the lower extreme part of the $\theta - v$ diagram (Figure 2) is now removed by the Monte Carlo simulation.

To our surprise, in some cases, the optimal $\theta_0$ and $v_0$ lie in an unexpected part of the $\theta - v$ diagram. For a direct shooting from $(2,3)\,m$ with a release height of $z = 1.9\,m$, the optimal launch condition obtained from the Monte Carlo simulation is located far down the horseshoe shape (Figure 6). We also observe that there exists a competition between high launch angle trajectories and low launch angle trajectories. At some point in the court, the competition may become close, increasing the variety and complexity of the basketball shooting.

It is noteworthy that the points in the $\theta - v$ diagram may occupy different phase space volumes. Thus, we need to trace all the successful shots in the $\vec{v}$-space. We opted to divide the phase space into cubic cells of each side $0.01\,m/s$ and count the number of occupied cells to estimate $\Omega$.

Although there exist no reliable data, we have an impression that smaller basketball players are more likely to utilize bank shots. Since deliberate bank shots are attempted closer to the hoop, we considered the court positions of the players $(x, y)$, where $x$ and $y$ are 1 $m$, 2 $m$, and 3 $m$, and calculated the phase space volume at each court positions for the taller player with $z = 2.1\,m$ and the smaller player with $z = 1.9\,m$.

The results for the bank shots are summarized in Table I and the results for the direct shots in Table II; $\Omega_{z=2.1} - \Omega_{z=1.9}$ for a shooter at each grid points in the court. The taller player does better, whether in bank shots or in direct shots, when he shoots from $(1,1)\,m$, that is, very near to the hoop. However, from approximately 1 $m$ to 3 $m$ range, the shorter player has an overall advantage in bank shots and the taller player in direct shots. These findings are consistent with our previous impression.

It is also of interest to see if bank shots can be strategically better choices, and if so, under what conditions. Phase space volume difference between the bank shots and the direct shots are summarized in Table III and Table IV. For the shorter player with $z = 1.9\,m$ (Table III), the bank shots had more phase space volume than the direct shots when the shooter was closer to the center line. However, the direct shots were favorable if the shooter was located near to the backboard and/or the sideline. This result is in agreement with those of Silverberg *et al* [5].

The taller player ($z = 2.1\,m$) again does not benefit from bank shots as much as the shorter player (Table IV). The difference in the phase space volume is significantly reduced even at those court positions where the bank shots are favored over the direct shots. These findings probably explain why those tall professional basketball players do not choose bank shots as often as shorter amateur players do.

## IV. CONCLUSION

We propose to view the basketball shootings as a Monte Carlo sequence in which a player judges his own shooting by how far he has missed the hoop. Then, it becomes possible to measure the difficulty of a shot by the associated phase space volume from Monte Carlo simulations. Our results offer some explanations why taller players less frequently opt for bank shots than shorter players and where in the court are more favorable positions to try bank shots without introducing *ad hoc* assumptions on the launch velocity distributions.

## ACKNOWLEDGMENTS

This research was supported in part by the Daegu University Research Funds.

Table I. Phase space volume difference $\Omega_{z=2.1} - \Omega_{z=1.9}$ for bank shots by a shooter at $(x, y)$ in units of $1 \times 10^{-3} (m/s)^3$.

|       | $x=1$ | $x=2$ | $x=3$ |
|-------|-------|-------|-------|
| $y=1$ | 7.3   | -5.1  | -0.1  |
| $y=2$ | -14.3 | -16.0 | 3.6   |
| $y=3$ | -25.3 | -0.9  | -10.3 |

Table II. Phase space volume difference $\Omega_{z=2.1} - \Omega_{z=1.9}$ for direct shots by a shooter at $(x, y)$ in units of $1 \times 10^{-3} (m/s)^3$.

|       | $x=1$ | $x=2$ | $x=3$ |
|-------|-------|-------|-------|
| $y=1$ | 6.6   | -4.4  | 0.4   |
| $y=2$ | -1.9  | 3.7   | 0.2   |
| $y=3$ | -1.4  | -0.8  | -0.2  |

Table III. Phase space volume difference $\Omega_{bank} - \Omega_{direct}$ for the shorter shooter ($z = 1.9\,m$) at $(x, y)$ in units of $1 \times 10^{-3} (m/s)^3$.

|  | $x = 1$ | $x = 2$ | $x = 3$ |
|---|---|---|---|
| $y = 1$ | -1.6 | -6.5 | -25.3 |
| $y = 2$ | 12.6 | 22.0 | -1.3 |
| $y = 3$ | 21.8 | 1.4 | 9.5 |

Table IV. Phase space volume difference $\Omega_{bank} - \Omega_{direct}$ for the taller shooter ($z = 2.1\,m$) at $(x, y)$ in units of $1 \times 10^{-3} (m/s)^3$.

|       | $x = 1$ | $x = 2$ | $x = 3$ |
|-------|---------|---------|---------|
| $y = 1$ | -0.8 | -7.2 | -3.0 |
| $y = 2$ | 0.3  | 2.3  | 2.1  |
| $y = 3$ | -2.1 | 1.2  | -0.7 |

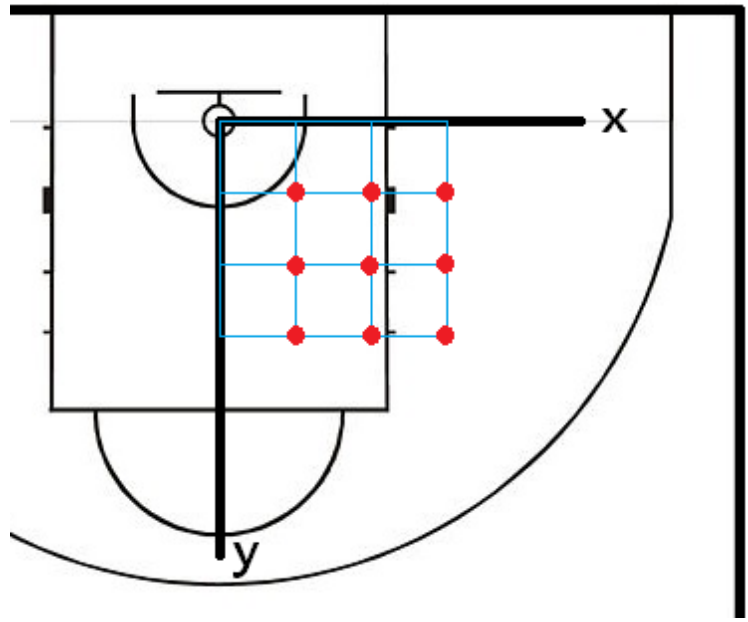

Fig.1

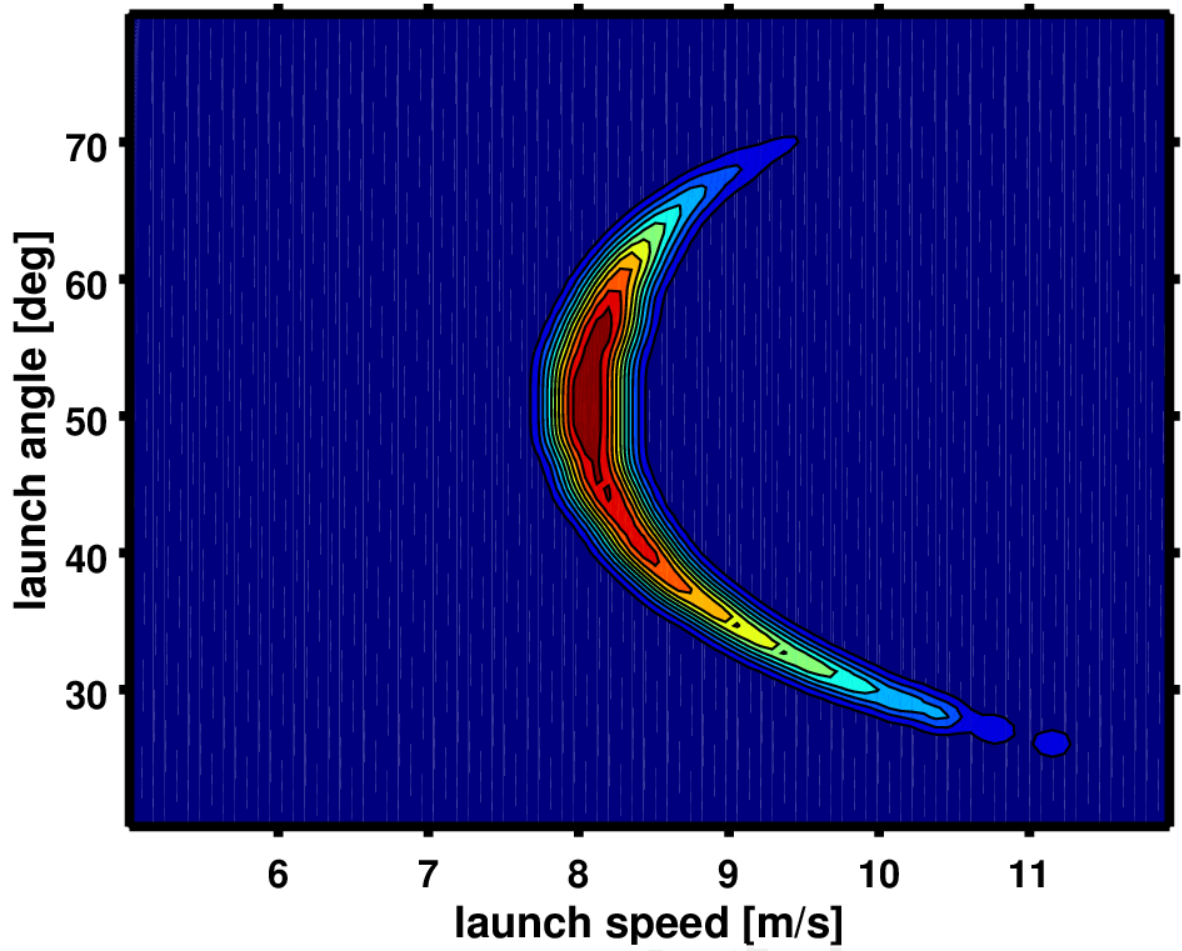

Fig.2

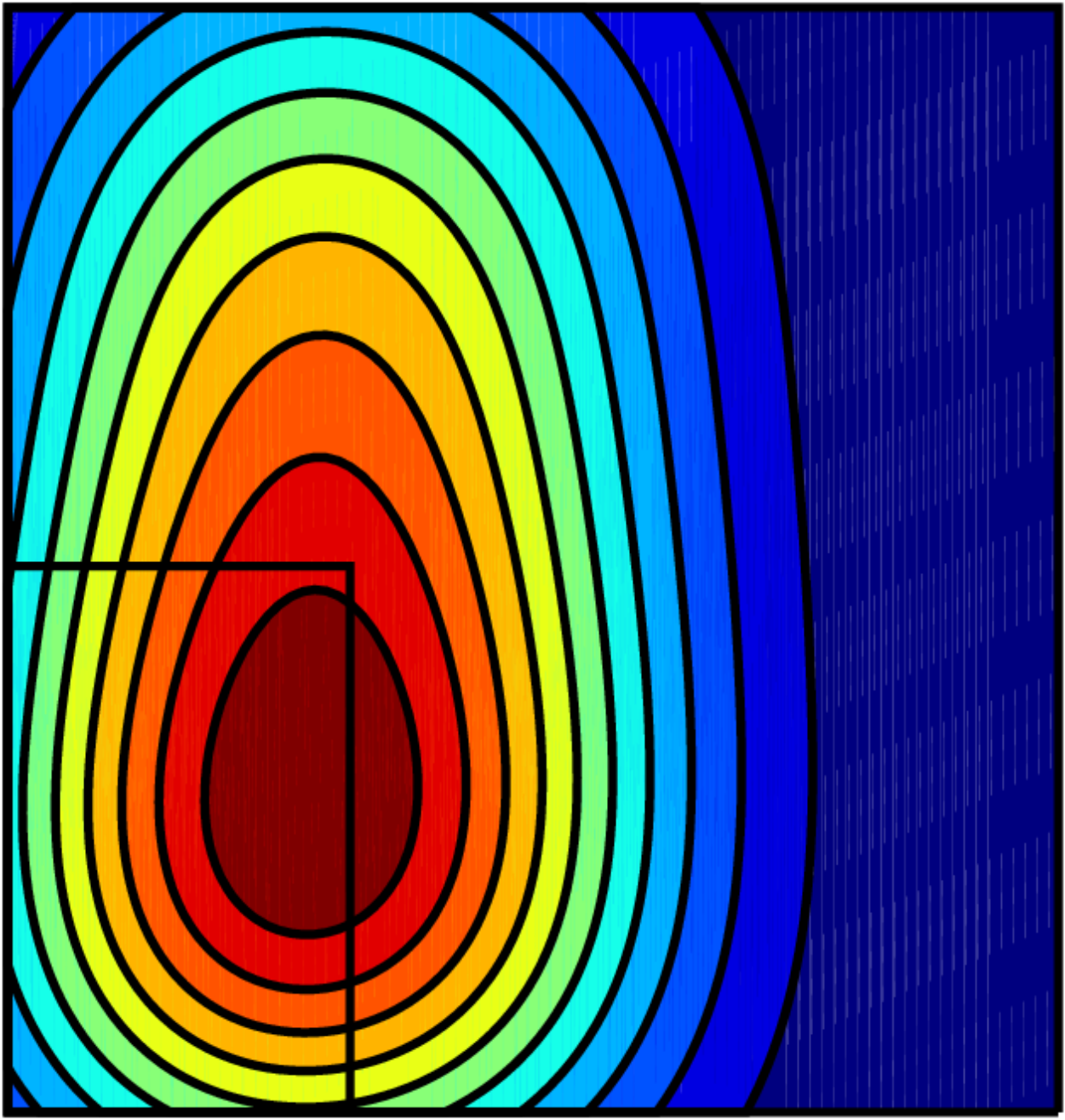

Fig.3

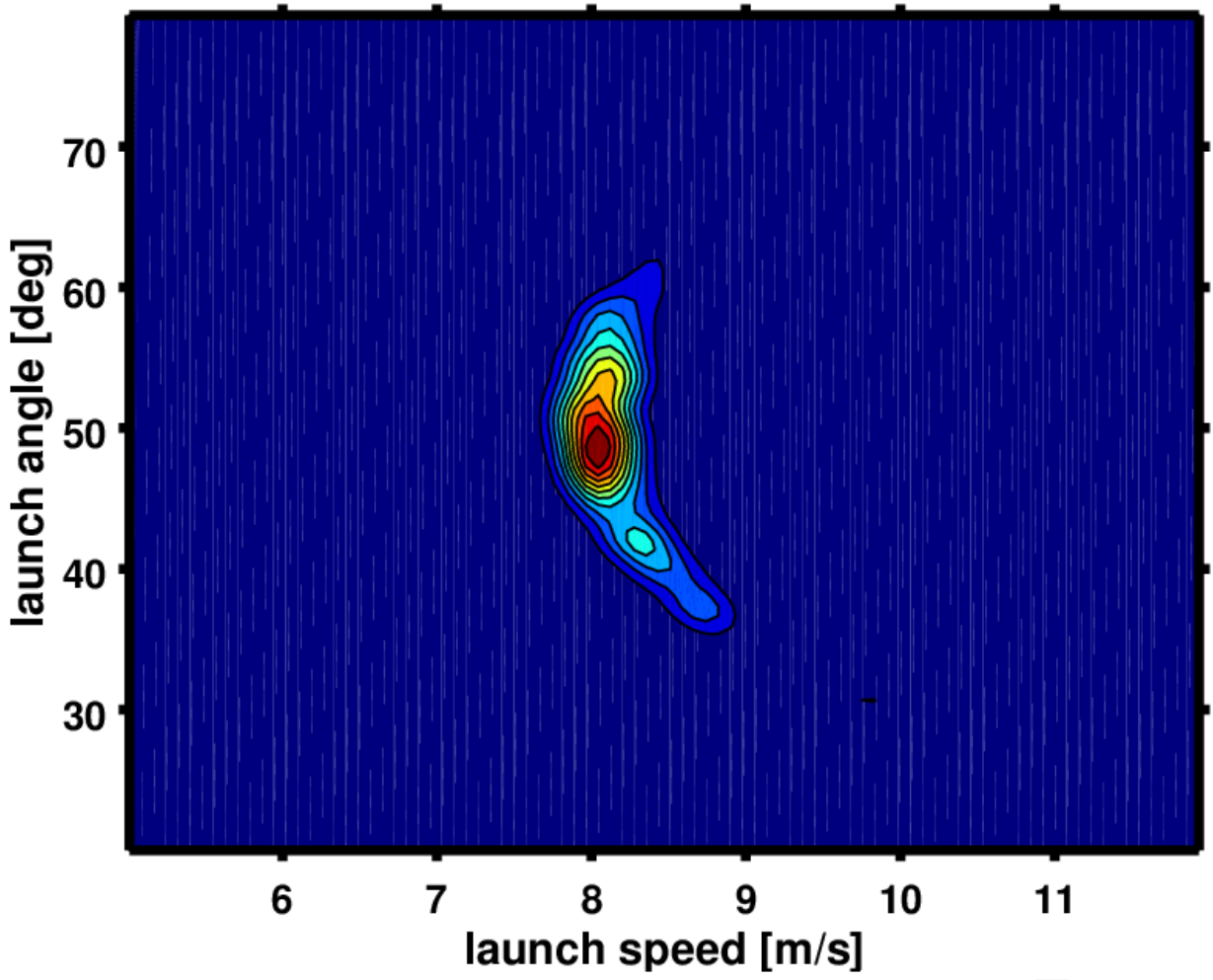

Fig.4

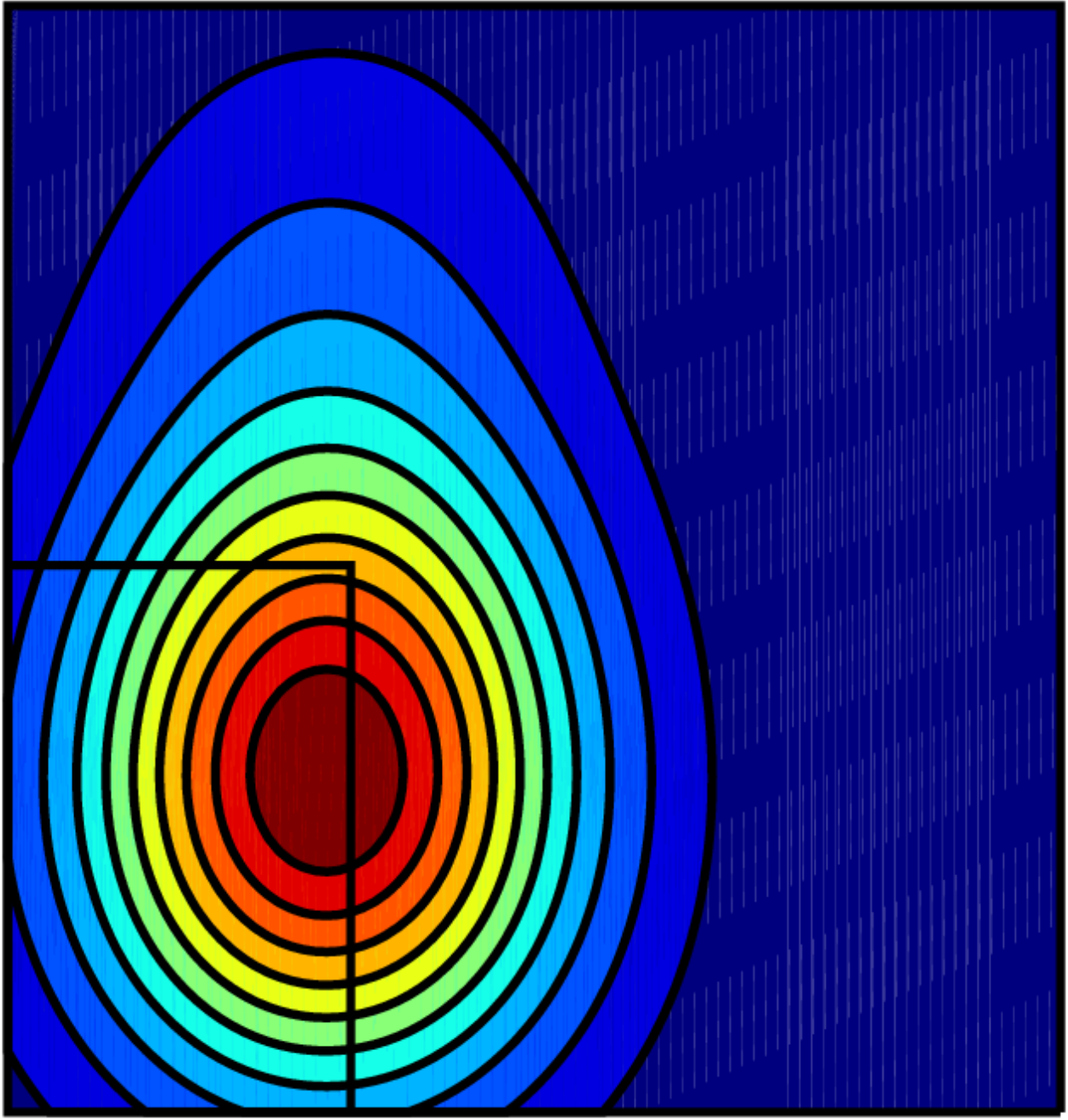

Fig.5

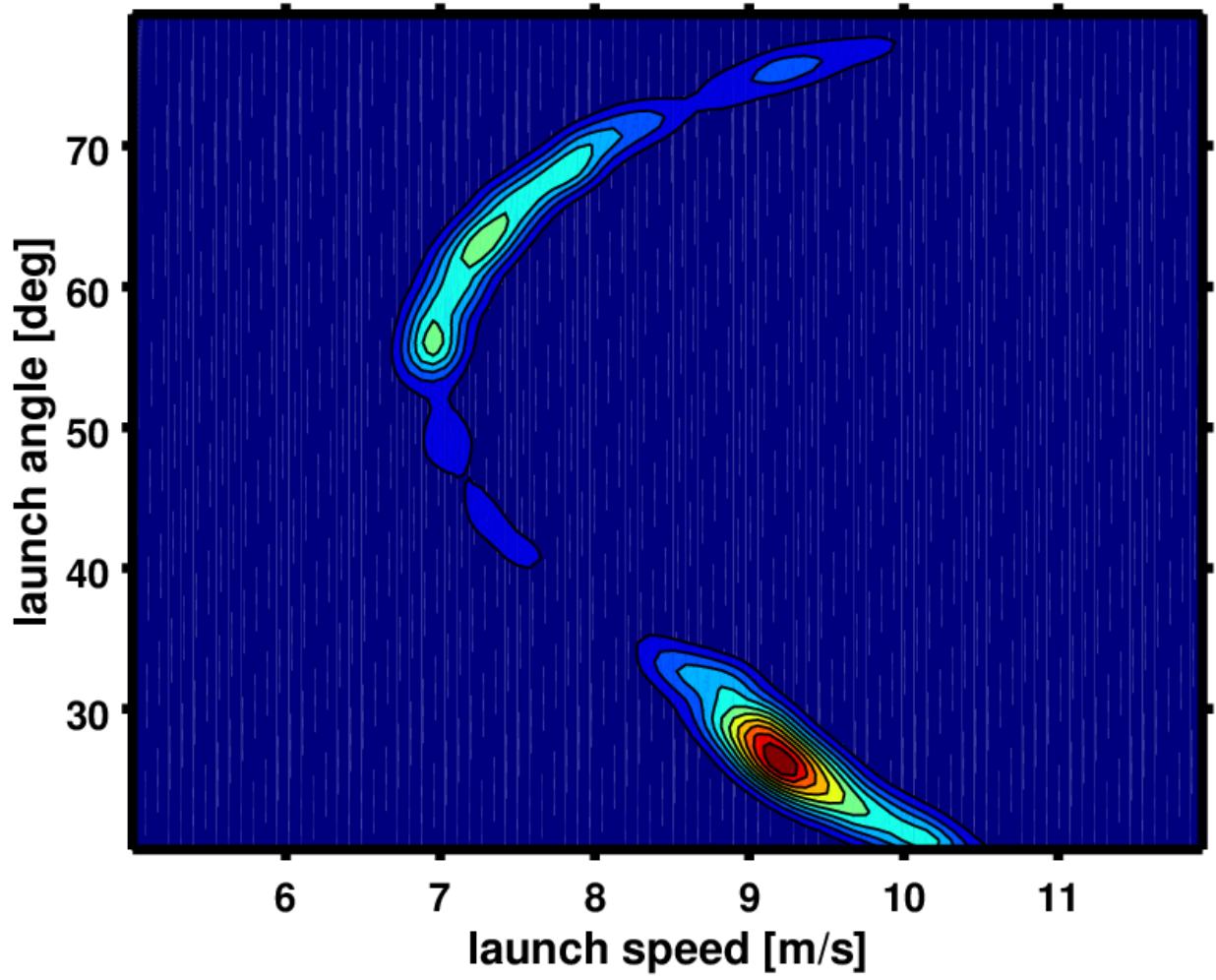

Fig.6

Figure Captions.

Fig. 1. The basketball court diagram near the hoop. We consider 9 positions near the hoop given by $(x, y)$, where $x, y = 1, 2, 3\, m$ as marked in the diagram.

Fig. 2. The launch angle vs. the launch speed diagram for all the bank shots that can be made by a shooter located at $(3,3)\, m$ from the center of the hoop with a release height of $1.9\, m$.

Fig. 3. The probability density distribution of bank shots in Figure 2 on the backboard. Only the right half of the backboard above the rim height is shown. The horizontal axis is the x-axis, and the vertical the z-axis.

Fig. 4. The launch angle vs. the launch speed diagram for the successful bank shots in the Monte Carlo sequence by a shooter located at $(3,3)\, m$ from the center of the hoop with a release height of $z = 1.9\, m$.

Fig. 5. The distribution of successful bank shots in the Monte Carlo sequence as described in Figure 4 on the backboard.

Fig. 6. The launch angle vs. the launch speed diagram for the successful direct shots in the Monte Carlo sequence by a shooter located at $(2,3)\, m$ from the center of the hoop with a release height of $z = 1.9\, m$. The optimal launch parameters are located far from the center of the horseshoe shape.